\documentclass[conference, 9pt]{IEEEtran}
\IEEEoverridecommandlockouts

\usepackage[utf8]{inputenc}
\usepackage{booktabs}
\usepackage{graphicx}
\usepackage{subfigure}
\usepackage{units}
\usepackage{blindtext}
\usepackage{xcolor,xspace}
\usepackage{amsmath}
\usepackage{paralist}
\usepackage{tikz}
\usepackage{booktabs}
\usepackage{pifont}
\usepackage{multirow}
\usepackage{hhline}
\usepackage{balance}
\usepackage{footnote}
\usepackage{logicproof}
\usepackage{caption}
\usepackage{amsthm}
\usetikzlibrary{patterns}
\usetikzlibrary{calc}
\usetikzlibrary{decorations.pathreplacing}
\usetikzlibrary{arrows,automata}
\usetikzlibrary{arrows,calc,shapes,decorations.pathreplacing}
\usepackage{pbox}
\usepackage{tabularx}
\usepackage{float}

\usepackage{arydshln}
\usepackage{mdframed}
\usepackage{pgf-umlsd}

\usepackage[T1]{fontenc}
\usepackage{beramono,textcomp}
\usepackage{enumitem}

\newcommand{\revised}[1]{\textcolor{black}{#1}}

\usepackage{listings}
\usepackage[ruled]{algorithm2e}
\usepackage{algorithmic}

\LinesNumbered

\usepackage{url}

\expandafter\def\csname ver@l3regex.sty\endcsname{}
\usepackage[n,advantage,operators,sets,adversary,landau,probability,notions,logic,ff,mm,primitives,events,complexity,asymptotics,keys]{cryptocode}
\usepackage{float}

\definecolor{mygreen}{rgb}{0,0.6,0}
\definecolor{mygray}{rgb}{0.5,0.5,0.5}
\definecolor{mymauve}{rgb}{0.58,0,0.82}

\newcommand{\fora}{\ensuremath{\sf{\mathcal PARseL}}\xspace} %
\newcommand{\foratext}{Verified Root-of-Trust over \sel \xspace} 

\newcommand{\foratextunderline}{\underline{P}rovable \underline{A}ttestation \underline{R}oot-of-Trust over {\sf \underline{seL}}4\xspace} 
\newcommand{\foratextshort}{Verified Root-of-Trust over \sel\xspace}

\newcommand{\Clang}{\texttt{\footnotesize C}\xspace}

\newcommand{\RA}{{\ensuremath{\sf{\mathcal RA}}}\xspace}

\renewcommand{\key}{\ensuremath{\mathcal K}\xspace}
\renewcommand{\prover}{{\ensuremath{\sf{\mathcal Prv}}}\xspace} 
\renewcommand{\verifier}{{\ensuremath{\sf{\mathcal Vrf}}}\xspace} 
\renewcommand{\adv}{\ensuremath{\sf{\mathcal Adv}}\xspace}

\newcommand{\hydra}{{\sf HYDRA}\xspace}
\newcommand{\vrased}{{{\sf\it VRASED}}\xspace}
\newcommand{\sel}{{\sf seL4}\xspace}
\newcommand{\hacl}{{\ensuremath{\sf HACL^*}}\xspace}
\newcommand{\fstar}{{\ensuremath{\sf\it F^*}}\xspace}
\newcommand{\lowstar}{{\ensuremath{\sf\it Low^*}}\xspace}
\newcommand{\karamel}{{\ensuremath{\sf\it KaRaMeL}}\xspace}

\newcommand{\dicestar}{{\ensuremath{\sf DICE^*}}\xspace}

\newcommand{\RP}{{\small \texttt{\footnotesize RP}}\xspace} 
\newcommand{\AP}{{\small \texttt{\footnotesize AP}}\xspace} 
\newcommand{\IP}{{\small \texttt{\footnotesize IUP}}\xspace} 
\newcommand{\PST}{{\small \texttt{\footnotesize PSMT}}\xspace} 
\newcommand{\SP}{{\small \texttt{\footnotesize SP}}\xspace} 
\newcommand{\mmap}{{\small \texttt{\footnotesize mmap}}\xspace} 
\newcommand{\UP}{{\small \texttt{\footnotesize UP}}\xspace} 

\long\def\ignore#1{}

\newcommand{\pid}{\ensuremath{\mathsf{P_{ID}}}\xspace}

\newcommand{\chal}{{\ensuremath{\sf{Chal}}}\xspace}

\mathchardef\mhyphen="2D

\usepackage{tikz}

\begin{document}

\title{\fora: Towards a \foratext}

\author{
\IEEEauthorblockN{Ivan De Oliveira Nunes}
\IEEEauthorblockA{\textit{RIT, USA} \\
ivanoliv@mail.rit.edu}
\and
\IEEEauthorblockN{Seoyeon Hwang}
\IEEEauthorblockA{\textit{UCI, USA} \\
seoyh1@uci.edu}
\and
\IEEEauthorblockN{Sashidhar Jakkamsetti}
\IEEEauthorblockA{\textit{UCI, USA} \\
sjakkams@uci.edu}
\and
\IEEEauthorblockN{Norrathep Rattanavipanon}
\IEEEauthorblockA{\textit{Prince of Songkla Univ., Thailand} \\
norrathep.r@phuket.psu.ac.th}
\and
\IEEEauthorblockN{Gene Tsudik}
\IEEEauthorblockA{\textit{UCI, USA} \\
gene.tsudik@uci.edu}
}


\maketitle

\begin{abstract} 
Widespread adoption and growing popularity of embedded/IoT/CPS devices 
make them attractive attack targets. On low-to-mid-range devices,
security features are typically few or none due to various constraints. Such devices 
are thus subject to malware-based compromise. One popular defensive measure is 
Remote Attestation (\RA) which allows a trusted entity to determine the current 
software integrity of an untrusted remote device. 

For higher-end devices, \RA is achievable via secure hardware components.
For low-end (bare metal) devices, minimalistic hybrid (hardware/software) \RA is
effective, which incurs some hardware modifications. That leaves certain mid-range
devices (e.g., ARM Cortex-A family) equipped with standard hardware components, e.g., a 
memory management unit (MMU) and perhaps a secure boot facility. In this space, \sel
(a verified microkernel with guaranteed process isolation) is a promising platform 
for attaining \RA. \hydra \cite{hydra} made a first step towards this, 
albeit without achieving any verifiability or provable guarantees.

This paper picks up where \hydra left off by constructing a \fora architecture, 
that separates all user-dependent components from the TCB. This leads to much 
stronger isolation guarantees, based on \sel alone, and facilitates formal verification. 
In \fora, We use formal verification to obtain several security properties for the isolated \RA TCB,
including: memory safety, functional correctness, and secret independence. We implement \fora in \fstar and specify/prove expected properties using  Hoare logic. Next, we automatically translate 
the \fstar implementation to \Clang using \karamel, 
which preserves verified properties of \fora \Clang implementation (atop \sel).
Finally, we instantiate and evaluate \fora on a commodity platform -- a SabreLite embedded device.
\end{abstract}


\section{Introduction}\label{sec:intro}
Internet-of-Things (IoT) and Cyber-Physical Systems (CPS) devices have become ubiquitous in 
modern life, including households, workplaces, factories, agriculture, vehicles, and public spaces. 
They often collect sensitive information and perform safety-critical tasks, such as monitoring 
vital signs in medical devices or controlling traffic lights.
Given their importance and popularity, these devices are attractive targets for attacks, 
such as the Colonial Pipeline attack in the American energy
grid~\cite{colonial_pipeline_attack_darkside} and Ukraine power grid 
hack~\cite{ukraine_power_grid_hack}. 

Attacks are generally conducted via software exploits and malware infestations that result
in device compromise. Remote Attestation (\RA) is a security service for detecting  
compromises on remote embedded devices. It allows a trusted entity (\verifier) to 
assess the software integrity of an untrusted remote embedded device (\prover). 
\RA serves as an important building block for other security services, such as proof of execution~\cite{apex,litehax}, control-flow and data-flow attestation~\cite{cflat,lofat,dialed,atrium,oat,geden2019hardware}, 
and secure software updates~\cite{asokan2018assured,casu}. 

Many prior \RA techniques (e.g., \cite{vrasedp,trustlite,tytan,sancus}) focused 
on low-end devices, that run one simple application atop ``bare metal''.
For example, SANCUS~\cite{sancus} is a pure hardware-based \RA architecture for low-end devices.
Whereas, 
\vrased~\cite{vrasedp} is a hybrid (hardware/software) \RA 
architecture, while 
PISTIS~\cite{pistis} is a software-only one.
All these architectures are unsuited for higher-end devices that execute multiple 
user space processes in virtual memory.

At the other end of the spectrum, enclaved execution systems~\cite{sgx,sanctum} implement 
\RA for user-level sub-processes (called enclaves) on high-end systems, e.g., desktops, 
laptops\revised{,} and cloud servers. However, they require substantial dedicated hardware support, 
thus making this approach unsuitable for the comparatively resource-constrained mid-range 
devices that we target in this work.

\hydra~\cite{hydra} is an \RA architecture aimed at \revised{such} mid-range 
devices. It does not require additional hardware support other than an (often present) 
memory management unit (MMU) and a secure boot facility. \hydra relies on a formally verified 
microkernel, \sel~\cite{klein2009sel4}, to provide strong inter-process memory isolation.
However, neither \hydra's implementation nor its integration with \sel, is formally verified.
Also, as discussed in Sections~\ref{subsec:bg_ra} and \ref{subsec:unverifiable_hydra},
\hydra implements both attestation and untrusted application-defined functionalities 
in the same runtime process. Thus, \hydra's trusted computing base (TCB) implementation is 
application-dependent\revised{,} and whenever an application changes, errors can be introduced within the TCB. 
As a consequence, even if the \RA component in \hydra were verified, application bugs could 
still undermine its security due to the lack of guaranteed isolation. 
Unfortunately, moving away from this model also introduces non-trivial architectural 
challenges (see Section~\ref{subsec:design_challenges}), requiring a clean-slate trust model.

Motivated by the above, this paper re-visits \hydra trust model and proposes \fora: 
\foratextunderline Microkernel
-- a design that separates user-dependent components from the \RA TCB. 
This new model addresses the aforementioned challenges, leading to proper isolation, 
and facilitates formal verification. Specifically, we use formal verification to prove 
security properties for the (now isolated) root-of-trust in \fora. Proven properties 
include memory safety, functional correctness, and secret independence. 
We then deploy and evaluate \fora verified \Clang implementation (atop \sel) 
on a commodity prototyping board, SabreLite~\cite{SabreLite}.
\fora implementation is publicly available at \cite{parsel_code_anon}. 

\noindent\textbf{Organization:} 
Section \ref{sec:bg} overviews background, followed by our goals and assumptions in 
Section \ref{sec:goals_assumptions}. \fora design is presented in Section \ref{sec:design} and its 
implementation details are in Section \ref{sec:implementation}, along with formal verification. 
\fora security analysis follows in Section~\ref{sec:security_analysis} and limitations are
discussed in Section~\ref{sec:limitations}. The paper concludes with the related work overview in 
Section~\ref{sec:related_work}.

\section{Background}\label{sec:bg}
This section provides background information on \sel, \RA, and formal verification tools. Given familiarity
with these topics, it can be skipped with no loss of continuity.

\subsection{\sel Microkernel~\cite{klein2009sel4}}\label{subsec:bg_sel4}
\sel is a member of the L4 family of microkernels. Functional correctness of its implementation, 
including the \Clang code translation~\cite{sewell2013translation}, is formally verified, 
i.e., the behavior of \sel \Clang implementation strictly adheres to its specification.
To provide provable memory isolation between processes, \sel implements a \textit{capability}-based access 
control model. A capability is an unforgeable token that represents a set of permissions that define 
what operations can be performed on the associated object at which privilege level.  
This enables fine-grained access control by granting or revoking specific permissions to individual 
components or threads. 
Also, user-space applications cannot directly access or modify their own capabilities, because each capability is 
stored in \textit{Capability Space} (CSpace) which is managed by \sel.
User applications interact with \sel through system calls and operate on their capabilities indirectly. 
Since \sel enforces strict access control and authorization checks for system calls, \sel retains the ultimate 
authority over capabilities and their allocation, revocation, and manipulation. 

As a micro-kernel, \sel provides minimal functionality to user-space applications. 
For example, inter-processes' data sharing requires the establishment of \textit{inter-process communication} 
(IPC) by invoking \textit{endpoint} objects, that act as general communication ports. 
Each \revised{endpoint} 
is given a capability by assigning it a unique identifier, called a ``\textit{badge}'', 
which identifies the sender process during communication. 
Each process is represented in \sel by its \textit{Thread Control Block} object which includes its associated CSpace and 
\textit{Virtual-address Space} (VSpace) and (optionally) an \textit{IPC buffer}. 
CSpace contains the capabilities owned by the process. VSpace represents the virtual memory space of the process, 
defining the mappings between virtual addresses (used by the process) and physical memory.
IPC buffer is a fixed region of memory reserved for IPC.
\revised{
To send or receive messages, a process places them in its \textit{message registers} which are put in the IPC buffer 
and then it invokes the capabilities within its CSpace via \sel system calls.
}

\subsection{\RA \& \hydra}\label{subsec:bg_ra}
As mentioned earlier, the goal of \RA is for a trusted \verifier to securely assess the software integrity 
of an untrusted \revised{remote \prover}. 
To do so, \verifier issues a unique challenge to \prover. Using the received challenge, \prover computes an authenticated measurement of its own software state. This measurement 
\revised{is} computed using either a \prover-\verifier shared secret or a 
\prover-unique private key for which \verifier knows the corresponding public key.
\prover returns the measurement to \verifier which authenticates it and decides on \prover's state (i.e., 
compromised or not).

To the best of our knowledge, the only relevant prior result that attempted to fuse \RA with \sel is \hydra \cite{hydra}.
It operates in three phases: \textit{Boot}, \textit{\sel Setup}, and \textit{Attestation}.
In Boot phase, \prover executes a ROM-resident secure boot procedure that verifies \sel binary. 
Upon verification, \prover loads all executables into RAM and passes control to the kernel.
In \sel Setup phase, the kernel sets up the user space and initializes the first process, 
\textit{attestation process} (\AP). The kernel then hands control to \AP after assigning all 
capabilities for all available memory locations to \AP and verifying \AP's binary.
\AP is then responsible for spawning all user processes with lower scheduling priorities and 
user-defined capabilities, initializing the network interface, and 
waiting for subsequent attestation requests. 
Finally, in Attestation phase (which comprises the rest of the runtime), 
upon receiving a \verifier-issued attestation request for a particular user-space process, 
\AP computes an HMAC~\cite{hmac} of the memory region of that process, using a symmetric key 
pre-shared with \verifier, and returns the result to \verifier. 

\hydra~\AP implements several system functions that are unrelated to \RA functionality. While this 
approach simplifies {\it Boot} and {\it \sel Setup} phases, it also makes \hydra verification challenging. 
We further discuss this in Section~\ref{subsec:unverifiable_hydra}.

\subsection{\fstar, \lowstar, and \karamel}\label{subsec:bg_fstar}
\fstar~\cite{fstar} is a general-purpose functional programming language with an effect system facilitating program verification.
Developers can write a program and its specifications in \fstar, representing that program's computational and side effects,
and then formally verify that it adheres to those specifications using automated theorem-proving techniques. 
The type system of \fstar includes dependent types, monad effects, refinement types, and the weakest precondition calculus,
which together allow describing precise and compact specifications for programs using Hoare logic~\cite{hoarelogic}. 
For example, \revised{Fig.}~\ref{fig:fstar_ex_ftn} shows two simple functions in \fstar.
While both take an integer as input and output its absolute value, {\texttt{\footnotesize  abs\_pos}} ``requires'' the 
input integer to be positive as \textbf{pre-condition} and ``ensures'' that the result equals the absolute value of 
{\texttt{\footnotesize  x}} as \textbf{post-condition}. The pre-condition of {\texttt{\footnotesize  abs\_pos}} 
can be instead written with refinement type input: {\texttt{\footnotesize (x : int \{x > 0\})}}.
Both have the {\texttt{\footnotesize  Pure}} effect, meaning that they are stateless functions, guaranteeing 
deterministic results and no side effects. {\texttt{\footnotesize  Tot}} is a special type of {\texttt{\footnotesize  Pure}}
with no pre-condition, i.e., it is defined for all possible values of input so that it terminates and returns an output.
 
\begin{figure}[h]
    \centering
    \begin{minipage}{0.7\linewidth}
    \begin{lstlisting}[language=C]
let abs (x : int) : Tot int 
  = if x >= 0 then x else -x

let abs_pos (x : int) : Pure int
(requires x > 0) (ensures fun y -> y = abs x) = x
    \end{lstlisting}
    \end{minipage}
    \caption{Example Functions in \fstar}\label{fig:fstar_ex_ftn}
\end{figure}

To support stateful programs, \fstar provides \texttt{\footnotesize ST} effect with the form:
\begin{center} 
{\scriptsize
	\texttt{ST (a:Type) (pre:s$\rightarrow$Type) (post:s$\rightarrow$a$\rightarrow$s$\rightarrow$Type)}
}
\end{center}
This means: for a given initial memory ``{\texttt{\footnotesize  h0:s}}'' that satisfies pre-condition 
``\texttt{\footnotesize (pre h0)} is true'', a computation ``{\texttt{\footnotesize  e}}'' of type 
``\texttt{\footnotesize  ST a (requires pre) (ensures post)}''
outputs a result ``{\texttt{\footnotesize  r:a}}'' and updates existing memory to final memory 
``{\texttt{\footnotesize h1:s}}'', which satisfies the  post-condition ``{\texttt{\footnotesize (post h0 r h1)}} is true''.

One notable feature of \fstar is \textit{machine integers} and arithmetic operations on them.
Machine integers model (un)signed integers with a fixed number of bits, e.g., \texttt{\footnotesize uint32} \revised{and} \texttt{\footnotesize int64}, \revised{while} \texttt{\footnotesize FStar.Int.Cast} module offers conversions between these types. 
Using machine integers ensures that input and computation result values fit in the given integer bit-width, \revised{preventing an unintentional arithmetic overflow}.
In addition, one can express their secrecy level, denoted by `\texttt{\footnotesize PUB}' or `\texttt{\footnotesize SEC}'. 
The former \revised{is} considered public and can be safely shared,
while the latter \revised{is} considered secret, i.e., \fstar guarantees no leaks for them. 
Specifically, it prevents information leakage from timing side-channels and clears all memory that 
contains \texttt{\footnotesize SEC}-level integers when they are no longer needed.

\lowstar~\cite{lowstar} is a subset of \fstar, targeting a carefully curated subset of \Clang features, 
such as the \Clang
memory model with stack- and heap-allocated arrays, machine integers, \Clang string literals, and a few system-level 
functions from the \Clang standard library. To support these features, \lowstar refines the 
memory model in \fstar by adding a distinguished set of regions modeling \Clang call stack -- so-called 
\textit{hyper-stack} memory model. For modeling \Clang stack-based memory management mechanism, \lowstar introduces 
a region called {\texttt{\footnotesize tip}} to represent the currently active stack frame and relevant operations, 
such as {\texttt{\footnotesize push}} and {\texttt{\footnotesize pop}}. \lowstar also introduces the 
{\texttt{\footnotesize Stack}} effect with the form below, to ensure that the stack tip remains unchanged 
after any pushed frame is popped and the final memory is the same as the initial memory:
\vspace*{0.2cm}\\ 
{\scriptsize    
    \hspace*{0.2cm} \texttt{\scriptsize Stack a pre post = 
	ST a (requires pre) (ensures \\
	\hspace*{1cm} ($\lambda$ h0 r h1 $\rightarrow$ post h0 r h1 $\wedge$ \\
	\hspace*{1cm} (tip h0 = tip h1) $\wedge$ ($\forall$ x. x $\in$ h1 $\Leftrightarrow$ x $\in$ h0)))}\vspace*{0.2cm}
 }

Programmers writing code in \lowstar can utilize the entire \fstar for proofs and specifications.
This is because proofs are erased at compile-time and only low-level \lowstar code is left and compiled to 
\Clang code.  Verified \lowstar programs can be efficiently extracted to readable and idiomatic 
\Clang code using the \karamel~\cite{karamel} compiler tool (previously known as \textit{KreMLin}).
\karamel implements a translation scheme from a formal model of \lowstar, $\lambda$ow$^*$, to CompCert Clight~\cite{clight}: 
a subset of \Clang. This translation preserves trace equivalence with respect to the original 
\fstar semantics.  Thus, it preserves the functional behavior of the program without side channels due to memory access 
patterns that could be introduced by the compiler. The resulting \Clang programs can be compiled with 
CompCert or other \Clang compilers (e.g., GCC, Clang).

\subsection{\hacl Cryptographic Library ~\cite{hacl}}\label{subsec:bg_hacl}
\hacl~\cite{hacl} is a formally verified cryptographic library written in \lowstar and compiled to readable 
\Clang using \karamel. Each cryptographic algorithm specification is derived from the published 
standard and covers a range of properties, including:
\begin{compactitem}
    \item \textit{Memory safety}: verified software never violates memory abstractions so that it is free from common 
    vulnerabilities due to reads/writes from/to invalid memory addresses, e.g., buffer overflow, null-pointer dereferences, 
    and use-after-free.
    \item \textit{Type safety}: software is well-typed and type-related operations are enforced, i.e., \hacl code respects interface, and 
    all the operations are performed on the correct types of data.     \item \textit{Functional correctness}: input/output of the software for each primitive conform to simple specifications 
    derived from published standards.
    \item \textit{Secret independence}: observations of the low-level behavior, such as execution time or accessed memory 
    addresses, are independent of secrets used in computation, i.e., the implementation is free of timing side-channels.
\end{compactitem}

\section{Goals \& Assumptions} \label{sec:goals_assumptions}
\subsection{System Model} \label{subsec:system_model}
We consider \prover to be a mid-range embedded device equipped with an MMU and a secure boot facility\footnote{Although common in mid-range embedded devices, secure boot requirement can be relaxed with weaker adversary model where \adv does not have physical access to \prover and the initial deployment of \sel and \fora TCB on \prover is trusted.}. 
Devices in this class include I.MX6 Sabre Lite~\cite{SabreLite} and HiFive Unleashed~\cite{HiFiveUnleashed} (on which \sel is fully formally verified \cite{sel4_supported_platforms}).
Following \sel verification axioms, \prover is limited to one active CPU core, i.e., it schedules multiple 
user-space processes, though only one process is active at a time.
We assume that secure boot is correctly enabled prior to device deployment. 

\fora TCB consists of \sel \revised{microkernel}, the first process loaded by the \revised{microkernel} in user-space, called \textit{Root Process} (\RP), and \textit{Signing Process} (\SP), also in user-space (details in Section~\ref{sec:design}).
\verifier wants to use \RA to establish a secure channel with a particular attested user-space process.
\revised{To facilitate this, \fora attestation response can also include a unique public key associated with the process.} 
\verifier can then use the secure channel to communicate sensitive data with the attested process, after verifying its integrity through \RA.

\fora provides a \textit{static} root of trust for measurement of user-space process, i.e., the binary of processes are measured at their loading time.
This is plausible because \fora, by design, enforces that no new user process is spawned during runtime and no modifications on code occur without rebooting the device.  
On the other hand, \fora design allows the user-process updates without modifying \fora TCB. 
However, any updates require the device to reboot to re-measure the updated programs, which limits the scalability. 
We further discuss this limitation and possible alternatives in Section~\ref{sec:limitations}.

\fora design is agnostic to the choice of cryptographic primitives.
In fact, \fora can support both (1) symmetric-key cryptography where \prover and \verifier share a master secret from which a subsequent symmetric key can be derived, or (2) public-key crypto-graphy where \prover has a private signing key whose public counterpart is securely provisioned to \verifier.
In both cases, the required keys can be hard-coded as part of the \fora TCB  prior to \prover deployment. 

\subsection{Adversary Model}\label{sec:adv_model}
Based on the \RA taxonomy in \cite{abera2016invited}, four main types of \adv are:
\begin{compactenum}
    \item \textit{Remote}: exploits vulnerabilities in \prover software and injects malware over the network;
    \item \textit{Local}: controls \prover's local communication channels; may attempt to learn secrets leveraging timing side-channels;
    \item \textit{Physical non-intrusive}: has physical access to \prover and attempts to overwrite its software through legal programming interfaces (e.g., via J-TAG or by replacing an SD card).
    \item \textit{Physical intrusive}: performs invasive physical attacks, physical memory extraction, firmware tampering, and invasive probing, e.g., via various physical side-channels.
\end{compactenum}
\revised{
We consider type (1) and (2) adversaries. Type (3) can be supported if \prover hardware offers protection to prevent access to \prover's secret key via programming interfaces.
Protection against type (4) adversaries is orthogonal and typically obtained via standard physical security measures~\cite{ravi2004tamper}.
}
This scope is in line with related work on trusted hardware architectures for 
embedded systems~\cite{smart,tytan,vrasedp,trustlite}. In terms of capabilities, if \adv compromises a user-space process in \prover, 
it takes full control of that user-space process, i.e., it can freely read and write its memory and diverge its control flow. 
We assume user-space processes as untrusted and therefore compromisable, except for \fora TCB.  
Finally, we assume that \adv can trigger interrupts at any time.

\begin{figure*}[h!tb]
    \centering
       \begin{subfigure}[\hydra Execution Levels]{ \centering
         \includegraphics[width=0.31\linewidth, keepaspectratio] {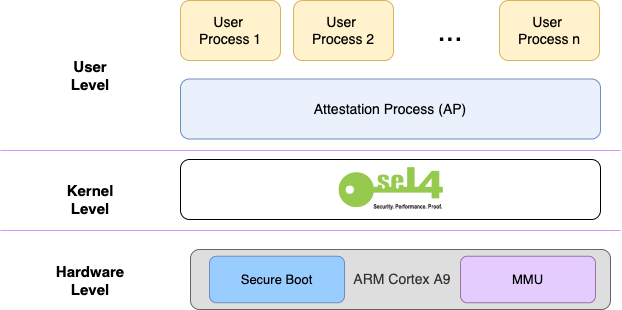}
         \label{subfig:hydra_exec_levels}
         }
       \end{subfigure}
       \hfill
       \begin{subfigure}[\fora on Boot]{ \centering
         \includegraphics[width=0.31\linewidth, keepaspectratio] {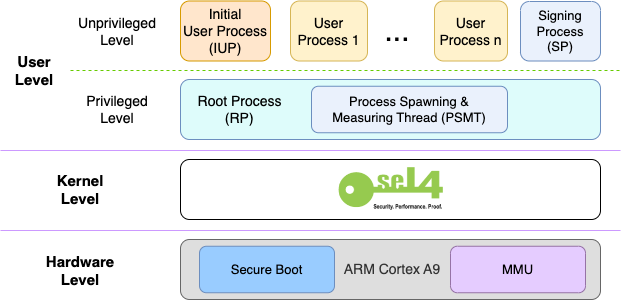}
         \label{subfig:fora_exec_levels_boot}
         }
        \end{subfigure}
        \hfill
       \begin{subfigure}[\fora at Runtime]{ \centering
         \includegraphics[width=0.31\linewidth, keepaspectratio] {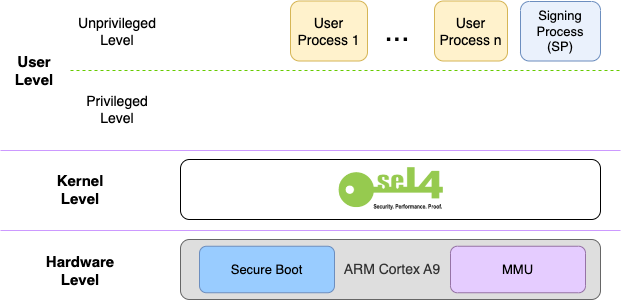}
         \label{subfig:fora_exec_levels_run}
         }
       \end{subfigure}
    \caption{Comparison of \hydra (left) and \fora Execution Levels on Boot (middle) and at Runtime (right)} \label{fig:exec_level_comparisons}
\end{figure*}

\section{\foratextshort (\fora)}\label{sec:design}
\revised{This section starts by describing \hydra and its limitations. }
It then justifies our approach and discusses how \fora realizes it.

\subsection{\hydra \& Its Limitations} \label{subsec:unverifiable_hydra}
As mentioned above, \hydra is composed of \textit{Boot}, \textit{\sel Setup}, and \textit{Attestation} phases.
\AP is the very first user-space process to run after \textit{\sel Setup}.
As such, \AP possesses all capabilities for all available memory and system resources. 
It is responsible for creating and managing all other processes, ensuring proper configuration 
of capabilities for them, and performing \RA.

We argue that this design results in an excessive and application-dependent TCB. First,
formally verifying the implementation of \AP is extremely challenging since it requires a
giant manual proof effort that might not be achievable in practice. However, without formal 
verification, there is no guarantee that \AP is vulnerability-free and correct.
Since \AP has all user-space capabilities, its compromise would lead to a breach of all \sel
isolation guarantees provided. Even assuming the feasibility of \AP formal verification,
process-spawning component of \AP strictly depends on the specific user application 
configuration.
This is so that \AP can properly assign custom (user-defined) access control configurations to 
each application process. Thus, whenever an application changes, \AP implementation needs to be 
adjusted accordingly. Doing so modifies the \AP's previously verified TCB. It is clearly 
infeasible to re-verify \AP implementation for all possible application-dependent configurations.

\subsection{Design Rationale} \label{subsec:design_challenges}
To enable verifiability, the TCB size at runtime must be reduced, by identifying and 
removing unnecessary functionalities from the privileged \AP process. \hydra~\AP 
functionalities are:
\begin{compactenum}
\item[\textcircled{\raisebox{-0.9pt}{1}}] Spawning all user processes with memory/capability settings;
\item[\textcircled{\raisebox{-0.9pt}{2}}] Communication with \verifier over the network interface for \RA;
\item[\textcircled{\raisebox{-0.9pt}{3}}] Attestation of all user processes; 
\end{compactenum}
First, we observe that including \textcircled{\raisebox{-0.9pt}{2}} in the TCB yields no benefit since 
the security of \RA does not depend on the availability/integrity of the communication interface. 
Thus, we move this 
functionality out of the TCB and handle \prover $\leftrightarrow$ \verifier communication in a separate user-space process.
Second, \textcircled{\raisebox{-0.9pt}{1}} performing initialization tasks that are not needed 
at runtime (i.e., post-boot). Third, further sub-dividing \textcircled{\raisebox{-0.9pt}{3}}:
\begin{compactenum}
    \item[\textcircled{\raisebox{-0.9pt}{3}}-(a)] Measuring (reading) the code binary for each user process;
    \item[\textcircled{\raisebox{-0.9pt}{3}}-(b)] Signing the measurement with a private key and a challenge from \verifier; 
\end{compactenum}
\textcircled{\raisebox{-0.9pt}{3}}-(a) can be also done once, assuming that the code does not change post-boot (as mentioned in Section \ref{subsec:system_model}).
Thus, these components can be terminated after completion, at boot time, which
effectively limits these components' exploitable time window to boot time.

Also, \textcircled{\raisebox{-0.9pt}{1}} can be sub-divided into:
\begin{compactenum}
    \item[\textcircled{\raisebox{-0.9pt}{1}}-(a)] Storing access control capabilities for all processes to be spawned;
    \item[\textcircled{\raisebox{-0.9pt}{1}}-(b)] Spawning the user processes based on given access control capabilities;   
\end{compactenum}
To separate all user-dependent components from the TCB, 
a separate user process can perform \textcircled{\raisebox{-0.9pt}{1}}-(a) and communicate with \AP for \textcircled{\raisebox{-0.9pt}{1}}-(b). Or it can be even just a configuration file that \AP can read from. 
Finally, \textcircled{\raisebox{-0.9pt}{3}}-(b) must be active at runtime to process \RA requests from \verifier, 
which represents the only potential remaining entry point for \adv.  
To close this gap, this operation can be assigned to a tiny dedicated process, called \emph{Signing Process} (\SP). 
Due to its small size and independence of user-defined components, verifying \SP is now relatively easier.

\subsection{\fora Design}
Combining all the above, Fig.~\ref{fig:exec_level_comparisons} shows \fora execution levels at boot- and at run-time, 
as compared to \hydra. \fora subdivides \sel user-space into two execution levels: \textit{Privileged} and 
\textit{Unprivileged}. 
We refer to the privileged initial user process as \textit{Root Process} (\RP) which 
has a thread (for the roles of \textcircled{\raisebox{-0.9pt}{1}}-(b) and \textcircled{\raisebox{-0.9pt}{3}}-(a)), 
called \textit{Process Spawning \& Measuring Thread} (\PST). 
In contrast, the processes at the unprivileged level have restricted capabilities assigned by \RP.  
Unprivileged processes include \textit{Initial User Process} (\IP) (for \textcircled{\raisebox{-0.9pt}{1}}-(a)), 
\SP, and user-defined processes (\UP-s). Capabilities of any process at the unprivileged 
level do not allow access to any memory not explicitly assigned to that process. 
\RP (including \PST) and \IP are terminated at the end of boot phase, 
and only \UP-s and \SP remain during run-time, as shown in Fig.~\ref{subfig:fora_exec_levels_run}.

\subsection{\fora Execution Phases}\label{subsec:execution_phases}

\begin{figure*}[h!tb]
	\centering
	\includegraphics[width=1\textwidth]{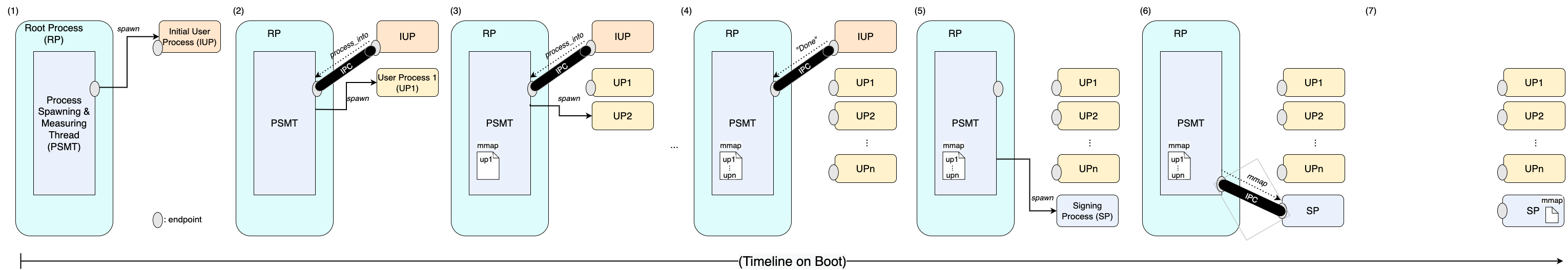}
	\caption{\footnotesize Sequence of \fora Execution Phases at Boot (After Secure Boot Checks)}
	\label{fig:execution_sequence_boot}
\end{figure*}

\begin{figure}
	\centering
	\includegraphics[width=0.49\columnwidth]{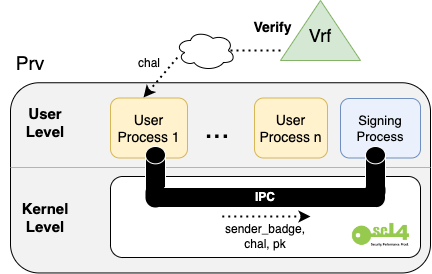}
         \includegraphics[width=0.49\columnwidth]{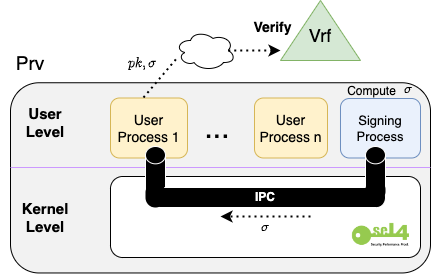}
	\caption{\footnotesize Sequence of \fora Execution Phases at Runtime}
	\label{fig:execution_sequence_runtime}
\end{figure}

\fora has seven execution phases in total: three on boot and four at runtime.
Three phases in the boot-time are:
\begin{description}
    \item[\textbf{(Secure) Boot}:] 
    The boot-loader verifies, loads, and passes control to, \sel.
    Thereafter, \sel verifies the integrity of \fora TCB, i.e., the software that runs in \RP, and passes control to \RP, once verification succeeds.
    \item[\textbf{Process Spawn}:] \RP spawns \PST as a thread. 
    \PST spawns \IP as an unprivileged process and establishes an IPC channel with it. 
    Once spawned, \IP sends the configuration of user processes and their process ID-s (\pid-s) to 
    \PST via IPC.
    Upon receiving a request, \PST spawns a new process according to received capabilities. It also ensures that these 
    capabilities are valid, not containing the write capability for its own code segment. 
    Finally, it spawns and sets up an IPC channel with \SP.
    Once all processes are spawned, \PST sets up an IPC \textit{endpoint} for each user process,
    assigns a unique \textit{badge} for each endpoint, 
    and associates this unique badge with \pid.
    \item[\textbf{Measurement}:] While spawning each user process, \PST also measures (via hashing) its code segment, and stores the results in \textit{measurement map} (\mmap) with the \pid as the lookup key.
    Once all measurements are complete, \PST sends the entire \mmap to \SP through IPC, and \RP 
    (including \PST) is terminated.
\end{description}
Once \prover is booted and in a steady state, it repeatedly executes the remaining four phases at runtime:
\begin{description} 
    \item[\textbf{Listen}:] \SP listens to receive messages from user processes through the endpoint 
    set up in the boot phase. 
    \item[\textbf{Request}:] Once a user process, \UP, receives an attestation request from \verifier 
    with a fresh challenge, \chal, \UP transmits the request to \SP through IPC system calls. 
    The request message includes \chal and the public key of \UP, \pk.  
    \item[\textbf{Sign}:] Upon receiving a request, \SP identifies the sender process, \UP, from the activated endpoint badge and derives \pid\footnote{Note that \sel guarantees that \UP cannot forge its own endpoint badge. Therefore, the attested \UP is the same process that provides \pk~to \SP.}. It then retrieves \UP's measurement $\texttt{m}_\UP$ 
    from \mmap using \pid and signs $\texttt{m}_\UP$ along with the request message using its 
    secret key, \key. i.e., 
    \begin{equation}\label{eq:sign}
        \sigma := Sign(\key, Hash(\chal || \pk || \texttt{m}_\UP)) 
    \end{equation}
    \item[\textbf{Response}:] \SP responds $\sigma$ to \UP via IPC. 
    \UP forwards $\sigma$ and \pk~to \verifier. Finally, after 
    successful $\sigma$ verification, \verifier establishes a secure channel with \UP 
    using the received \pk.
    \ignore{
    Verification involves checking whether 
    \sigma$    Sign(\key, Hash(\chal || \pk || \texttt{M}_\UP))
    $ 
    if public key cryptography is used, or: 
    $
        Vrf(\tilde{\key},\sigma,Hash(\chal || \pk || \texttt{M}_\UP)) = 1
    $ 

    \noindent where $\texttt{M}_\UP$ is \UP measurement computed by \verifier, and 
    $\tilde{\key}$ is the public component of \key in the asymmetric (\textit{Sign,Vrf}) scheme.
}
\end{description}
Fig.s~\ref{fig:execution_sequence_boot} and \ref{fig:execution_sequence_runtime} 
show the aforementioned \fora execution phases on boot and at run-time, respectively.

\section{\fora Implementation \& Verification}\label{sec:implementation}

\subsection{Implementation Details} \label{subsec:sel4_impl}

\subsubsection{Implementation of \RP}
Once \sel passes control to \RP, 
\RP initializes user space by creating necessary boot-time objects, such as CSpace, VSpace, and a memory allocator.
Then, it initializes \PST by creating a new thread control block object, a memory frame for its IPC buffer, a new page table, and a new endpoint object.
Next, \RP maps the page table and IPC buffer frame into the VSpace and
configures a badge for the endpoint and thread control block priority. 
\RP then sets up the thread-local storage (for its own storage area) and spawns \PST. Finally, it waits for \PST to complete and send ACK.

\subsubsection{Implementation of \PST}
Once spawned, \PST creates \SP by assigning it a new set of virtual memory, configuring it with two endpoints, and associating a unique badge for each endpoint. 
\SP uses one endpoint for IPC with \SP and the other for \UP-s.
\PST similarly creates \IP, establishes an IPC between itself and \IP, and spawns \IP. 
Then, \PST waits for a request from \IP.

A request includes all the specifications of \UP to be spawned, such as \pid, binary location, and capabilities to system resources.
Once receiving the request, \PST first ensures that the requested capabilities do not contain the write capability to \UP's binary and then initializes \UP accordingly. 
Next, \PST computes its measurement, using a hash algorithm (e.g., SHA2-256~\cite{sha256}) in \hacl, and stores it in \mmap in order. 
\PST uses a counter to make sure the number of spawned processes does not exceed the size of \mmap and assigns a badge based on the counter to make it unique per \UP endpoint.
Finally, \PST spawns \UP and waits for the next request.
Once receiving the ``Done'' signal from \IP, \PST sends the entire \mmap to \SP via IPC, waits for \IP to finish its tasks (if any), and sends an ACK to \RP.

\subsubsection{Implementation of \IP}
In \fora, all the user process information is consolidated into a configuration file at compile-time. 
\IP first parses this file and loads its information to a local object.
Then, for each \UP, \IP sends a spawn request to \PST with its \pid, and waits for an acknowledgment.
After all the \UP-s are spawned, \IP sends the ``Done'' signal to \PST and finishes its remaining tasks (if any), before terminating itself.
Note that if \IP contains no tasks other than requesting to spawn \UP-s, then \PST can directly read the configuration file and spawn \revised{\UP-s}, 
instead of having a separated \IP.

\subsubsection{Implementation of \SP}
\SP has two roles: (1) collecting all the \UP-s measurements from \PST at boot-time, and (2) repeatedly processing \RA requests at runtime. 
Once \SP is spawned by \PST during boot-time, \SP uses \sel system calls to receive the entire \mmap via IPC in the following way: 
\begin{compactenum}
    \item Using \texttt{\footnotesize seL4\_Recv()}, \SP listens for measurement message (\pid, \texttt{m}) from \PST's badge.
    \item \SP uses \texttt{\footnotesize seL4\_GetMR()} to unmarshal the message and copies (\pid, \texttt{m}) to \mmap.
    \item Using \texttt{\footnotesize seL4\_Reply()}, \SP sends '0' (as a ACK).
\end{compactenum}
This process is repeated until all the measurements are received from \PST.
In the following section, we describe the verified implementation of \SP's runtime phase.

\subsection{Formally Verification of \fora Runtime Implementation} \label{subsec:fstar_impl}

We describe the implementation of runtime \fora TCB in \lowstar, with verified properties, and how to convert it to \Clang code, preserving the verified properties, using \karamel.

\subsubsection{Verifying Properties}

Recall that \SP runs the infinite loop of (\textit{Listen, Request, Sign, Response}) phases (see Section~\ref{subsec:execution_phases}). 
To verify \SP, we prove the following invariant properties for this infinite loop:
\textbf{functional correctness, memory safety}, and \textbf{secret independence}. 

\textit{Functional correctness} ensures that each loop iteration performs all the functionalities as intended.
In this context, it means each iteration of \SP correctly computes the signature according to Equation (\ref{eq:sign}) for the given input and returns the computed result without modifying \SP internal states.
\textit{Memory safety} and \textit{secret independence} guarantee that no additional information beyond the signature result is leaked from \SP. 
This applies to both memory-based leakages as well as timing side channels.
In Section~\ref{sec:security_analysis}, we show that these three properties are sufficient to provide secure \RA in \fora.

\subsubsection{Runtime \SP Implementation in \lowstar and \Clang}

To prove these properties, we first specify all \sel APIs used by \SP in \lowstar. Then, we implement the \lowstar code for all \SP execution phases and integrate it with the \lowstar-specified \sel APIs and \hacl verified cryptographic functions. Next, we formally verify the combined implementation via \lowstar memory model, intermediate assertions, and post-condition of the \SP execution. Finally, we convert the final \lowstar code to \Clang using the verified \karamel compiler. \vspace{0.2cm}

\noindent\textbf{[Specifying \sel APIs in \lowstar{}]} While \SP is implemented in \lowstar, the functional correctness of \sel implementation (including system calls) is verified with a different formal specification language called Isabelle/HOL~\cite{isabelle}. 
Hence, we represent them as axioms, using the construct `\texttt{\footnotesize assume val}' in \fstar.
\fstar type checker accepts the given assumption without attempting to verify it, and these axioms are converted to `\texttt{\footnotesize extern}' in the generated \Clang code.
We specify the input/output of each \sel system call with required type definitions.

For example, Fig.~\ref{fig:sel4_GetMR} shows in order, the original \Clang code for a system call, \texttt{\footnotesize seL4\_GetMR} from \sel APIs, corresponding \lowstar implementation as an axiom, and the generated \Clang code using \karamel.
\texttt{\footnotesize seL4\_GetMR} has an integer input \texttt{\footnotesize i} and simply outputs the \texttt{\footnotesize i}-th element of \texttt{\footnotesize msg} array in \texttt{\footnotesize seL4\_IPCBuffer} with type \texttt{\footnotesize seL4\_Word}. 
Including the new type \texttt{\footnotesize seL4\_Word} for \texttt{\footnotesize uint64}\footnote{It is defined either \texttt{\scriptsize uint32} or \texttt{\scriptsize uint64} depending on the underlying architecture, and the example code is shown with \texttt{\scriptsize uint64 seL4\_Word}.}, all the definitions or structs in \sel (lines 1-12 of original \Clang code) are properly converted into \lowstar (lines 1-17 of \lowstar code).
Note that since there is no concept of the \textit{global variable} in functional programming, all global variables or structs used in \SP are represented in \texttt{\footnotesize state} type (lines 5-7 of \lowstar code), initialized in \texttt{\footnotesize st\_var} (lines 8-15) and defined in function \texttt{\footnotesize st} (lines 16-17). 
Once the \lowstar axiom is compiled with \karamel, generated \Clang code only contains one line of declaration (line 12 of generated \Clang code) without implementation.
The rest of \sel system calls used in \SP, \texttt{\footnotesize seL4\_Recv}, \texttt{\footnotesize seL4\_Reply}, and \texttt{\footnotesize seL4\_SetMR}, are similarly written as axioms.  \vspace{0.2cm}

\begin{figure}
    \centering
    \begin{minipage}{0.9\columnwidth}
    \begin{lstlisting}[language=C]
#define _seL4_int64_type    long long int
typedef unsigned _seL4_int64_type   seL4_Uint64;
typedef seL4_Uint64 seL4_Word; 
typedef struct seL4_IPCBuffer_ {
    seL4_Word msg[seL4_MsgMaxLength]; // seL4_MsgMaxLength = 120
} seL4_IPCBuffer __attribute__((__aligned__(sizeof(struct seL4_IPCBuffer_))));
extern __thread seL4_IPCBuffer *__sel4_ipc_buffer;
__thread __attribute__((weak)) seL4_IPCBuffer *__sel4_ipc_buffer;
LIBSEL4_INLINE_FUNC seL4_IPCBuffer *seL4_GetIPCBuffer(void) 
{
    return __sel4_ipc_buffer;
}
LIBSEL4_INLINE_FUNC seL4_Word seL4_GetMR(int i)
{
   return seL4_GetIPCBuffer()->msg[i];
}
    \end{lstlisting} \vspace*{-0.5em}
    \end{minipage}
    \begin{minipage}{0.9\columnwidth}
    \begin{lstlisting}[language=C]
type seL4_Word = uint64 
noeq type seL4_IPCBuffer = {
  msg : mbuffer seL4_Word 120; 
}
noeq type state = {
  ipc_buffer: ipc:seL4_IPCBuffer;
}
let st_var: state = 
  let msg = B.gcmalloc HS.root (I.u64 0) 120ul in 
  let ipc_buffer = {
    msg = msg;
  } in
  {
    ipc_buffer = ipc_buffer;
  }
val st (_:unit):state
let st _ = st_var
assume val seL4_GetMR 
( i : size_t )
: Stack seL4_Word 
( requires fun h0 -> (size_v i < 120) /\ (size_v i >= 0) ) 
( ensures fun h0 a h1 -> B.(modifies loc_none h0 h1) /\ a == B.get h1 (st ()).ipc_buffer.msg (v i))   
	\end{lstlisting} \vspace*{-0.5em}
	\end{minipage}
 \begin{minipage}{0.9\columnwidth}
    \begin{lstlisting}[language=C]
typedef uint64_t seL4_Word;
typedef uint64_t *seL4_IPCBuffer;
typedef struct state_s
{
  uint64_t *ipc_buffer;
} state;
state st_var;
state st()
{
  return st_var;
}
extern uint64_t seL4_GetMR(uint32_t i);
    \end{lstlisting}  \vspace*{-0.5em}
    \end{minipage}
    \caption{Simplified example \sel API in original \sel library (top), axiom in \fstar (middle), and generated header file in \Clang (bottom) }
    \label{fig:sel4_GetMR}
\end{figure}

\noindent\textbf{[Writing \SP in \lowstar, combining \hacl library]} 
The \textit{Sign} phase is implemented using cryptographic operations in \hacl which is also implemented in \lowstar and formally verified according to their specification. 
Thus, three \hacl functions for concatenation, hash, and sign, are integrated into one signing function for Equation~(\ref{eq:sign}). 
We use HMAC~\cite{hmac} for the symmetric signing algorithm with SHA2-256~\cite{sha256} hash function and EdDSA~\cite{EdDSA} for the asymmetric one.
Runtime \SP with the four execution phases is implemented by combining this signing function and the \sel axioms.

First, to receive/send a message through the IPC buffer or store intermediate computation results, we need some local \Clang arrays in \lowstar. 
For representing \Clang arrays, \lowstar provides the \texttt{\footnotesize Buffer} module.
In \lowstar, a buffer is a reference to a sequence of memory with a starting \texttt{\footnotesize index} and a \texttt{\footnotesize length}. 
We use \texttt{\footnotesize alloca} (or \texttt{\footnotesize create} from \hacl) for stack allocation, and retrieve/update the buffer contents using 
\texttt{\footnotesize index}/\texttt{\footnotesize upd} with the proper indices. 

Then, since the \textit{Sign} phase is in between two \sel system calls for \textit{Request} and \textit{Response} phases, proper type conversions are required. 
Specifically, \sel system calls use the type \texttt{\footnotesize seL4\_Word} and \hacl functions require the \texttt{\footnotesize uint8} input type. 
To safely convert back and forth between \texttt{\footnotesize uint8} buffer and \texttt{\footnotesize seL4\_Word} buffer (with big-endian), we use \texttt{\footnotesize uints\_to\_bytes\_be} and \texttt{\footnotesize uints\_from\_bytes\_be} of the \texttt{\footnotesize Lib.ByteBuffer} module in \hacl.  \vspace{0.2cm}

\noindent\textbf{[Formal Verification]}
To verify the \textit{functional correctness} of runtime \SP, we first specify necessary pre-/post-conditions for each \sel axiom.
For example, the \lowstar code in Fig.~\ref{fig:sel4_GetMR} shows that the function \texttt{\footnotesize seL4\_GetMR} correctly returns with the \texttt{\footnotesize i}-th element of \texttt{\footnotesize msg} array in \texttt{\footnotesize seL4\_IPCBuffer} (line 22).
Also, some properties are needed to be specified to verify that \SP internal states are not modified. In Fig.~\ref{fig:sel4_GetMR}, the post-condition \texttt{\footnotesize B.(modifies loc\_none h0 h1)} indicates that no locations are modified from \texttt{\footnotesize seL4\_GetMR} function call (line 22).

Next, we insert an assertion detailed in Fig.~\ref{fig:functional_correctness_sign_assertion} after the \textit{Sign} phase to ensure the functional correctness of the signing function, i.e., it correctly computes the signature according to Equation (\ref{eq:sign}).

\begin{figure}[h]
\centering
 \begin{minipage}{0.9\columnwidth}
    \begin{lstlisting}[language=C]
// h0 is the initial memory state and h1 is the state right after the signing function call, using ST.get ()
assert ( B.as_seq h1 sign_result_u8 == 
            Spec.Ed25519.sign (B.as_seq h0 s.sign_key) 
                (Spec.Agile.Hash.hash alg 
                    (Lib.Sequence.concat #uint8 #64 #32 
                        (Lib.Sequence.concat #uint8 #32 #32 
                        (B.as_seq h chal) (B.as_seq h pk)) 
                    (B.as_seq h measurement_process)) 
                )  
        );
    \end{lstlisting}  \vspace*{-0.5em}
    \end{minipage}
    \caption{Assertion for Functional Correctness of \textit{Sign}, equation~(\ref{eq:sign})}
    \label{fig:functional_correctness_sign_assertion}
\end{figure}

Finally, we check the invariance of \key and \mmap throughout the \SP execution via intermediate assertions and the post-condition of the runtime \SP function. 
Similar to the assertion above, it 
compares the \key and \mmap contents in the memory (\texttt{\footnotesize h}) after executing each function call with the ones in the initial memory (\texttt{\footnotesize h0}), specified in Fig. \ref{fig:invariance}.
This invariance along with the post-conditions of \sel APIs and the assertion in Fig. \ref{fig:functional_correctness_sign_assertion} implies the functional correctness of runtime \SP.

\begin{figure}[h]
\centering
 \begin{minipage}{0.9\columnwidth}
    \begin{lstlisting}[language=C]
assert (B.as_seq h0 s.mmap == B.as_seq h s.mmap);
assert (B.as_seq h0 s.sign_key == B.as_seq h s.sign_key);
    \end{lstlisting}  \vspace*{-0.5em}
    \end{minipage}
    \caption{Assertion for \key and \mmap invariance}
    \label{fig:invariance}
\end{figure}

For \textit{memory safety}, we first implement all \SP components with \texttt{\footnotesize Stack} effect, which prevents any memory leakage due to deallocated heap regions. 
We also check the ``liveness'' and ``disjointness'' of all buffers before they are referenced (via \texttt{\footnotesize live} and \texttt{\footnotesize disjoint} clauses), which prevents stack-based memory corruption.
The former guarantees that a buffer must be properly initialized and not de-allocated (so ``live") before it is used, whereas the latter ensures that all buffers used in \SP are located in separate memory regions without any overlap. 
Lastly, we specify a post-condition for every function in \SP to ensure that it modifies only the intended memory region. 
This can be done through the \texttt{\footnotesize modifies} clause with the form of \texttt{\footnotesize modifies s h0 h1}, which ensures that the memory \texttt{\footnotesize h1} after the function call may differ from the initial memory \texttt{\footnotesize h0} (before the function call) \textit{at most} regions in \texttt{\footnotesize s}, i.e., no regions outside of \texttt{\footnotesize s} are modified by the function execution.
For example, in Fig.~\ref{fig:sel4_GetMR}, \texttt{\footnotesize seL4\_GetMR} function ensures not to modify any memory location (with `\texttt{\footnotesize loc\_none}') in its post-condition (line 22).

Finally, for the \textit{secret independence}, we use the same technique employed by \hacl.
We use the secret machine integers for private values (i.e., \key), including all intermediate values, 
and do not use any branch on those secret integers. 
This ensures that the execution time or the accessing memory addresses are independent of the secret values so that the implementation is timing side-channel resistant.  \vspace{0.2cm}

\noindent\textbf{[Generating \Clang code using \karamel]} 
Finally, we carefully write a build system and generate readable \Clang code from our verified \lowstar code using \karamel.
It takes an \fstar program, erases all the proofs, and rewrites the program from an expression language to a statement language, performing optimizations. 
If the resulting code contains only \lowstar code with no closures, recursive data types, or implicit allocations, then \karamel proceeds with a translation to \Clang.

\karamel generates a readable \Clang library, preserving names so that one not familiar with \fstar can review the generated code before integrating it into a larger codebase. 
For example, 
the refinement type ({\texttt{\footnotesize b: B.buffer uint32 {B.length b = n}}}) in \lowstar is compiled to a \Clang declaration ({\texttt{\footnotesize uint32\_t b[n]}}), while referred to via ({\texttt{\footnotesize uint32\_t *}}) as \Clang pointer.

\ignore{
\karamel translates \lowstar to CompCert's Clight, a subset of \Clang, that can be compiled by classic \Clang compilers such as GCC, Clang, and Microsoft's \Clang compiler or CompCert.
Generated \Clang files have no dependencies beyond a single header file and C11 standard headers so that the \karamel output can be easily integrated into an existing source tree.
Note that this translation preserves the semantics and event traces of a program.
As in \hacl, we consider the problem of compiling \Clang to the verified assembly out of scope.
}

\subsection{Secure Boot of \sel and \fora TCB} \label{subsec:secure_boot}

Similar to \hydra, \fora relies on a secure boot feature to protect against a physical \adv attempting to re-program \sel and \fora TCB when \prover is offline. 
In \hydra, this feature works by having a ROM boot-loader validate \sel authenticity before loading it. Once \sel is running, it authenticates the user-space TCB by comparing it to a benign hash value, hard-coded within the \sel binary.
Since \hydra TCB is user-dependent, updating a user application implies a software update not only to the TCB but also to the \sel binary that stores the TCB referenced hash value, which can be inconvenient in practice.
Conversely, \fora TCB is user-independent, allowing user applications to be updated directly without the need to modify \fora TCB or \sel binary.

\begin{figure}
	\centering
	\includegraphics[width=0.49\columnwidth]{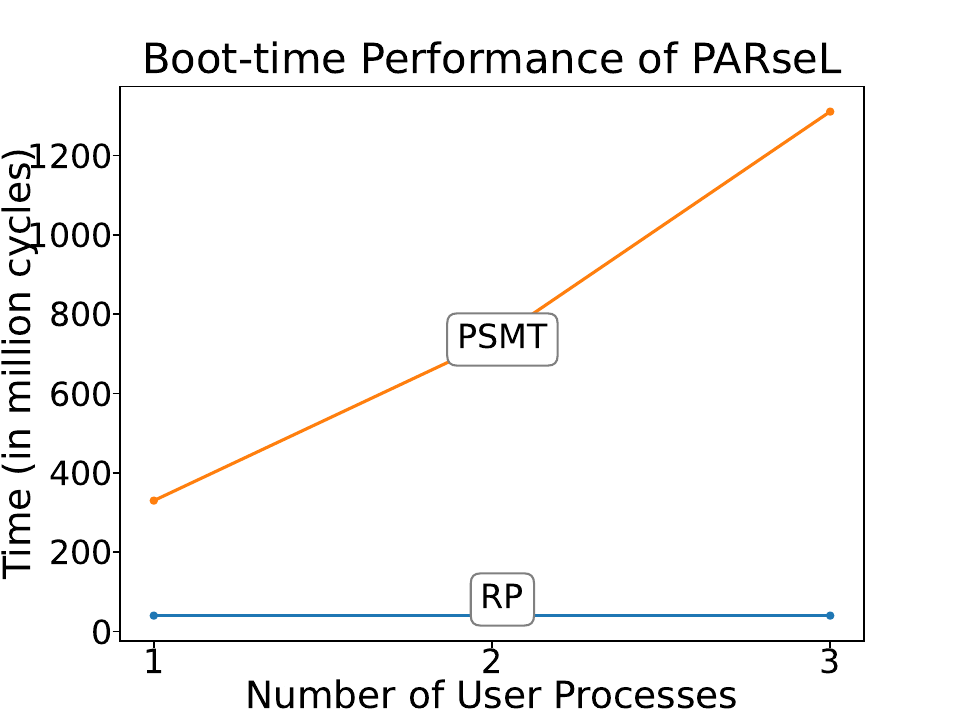}
        \includegraphics[width=0.49\columnwidth]{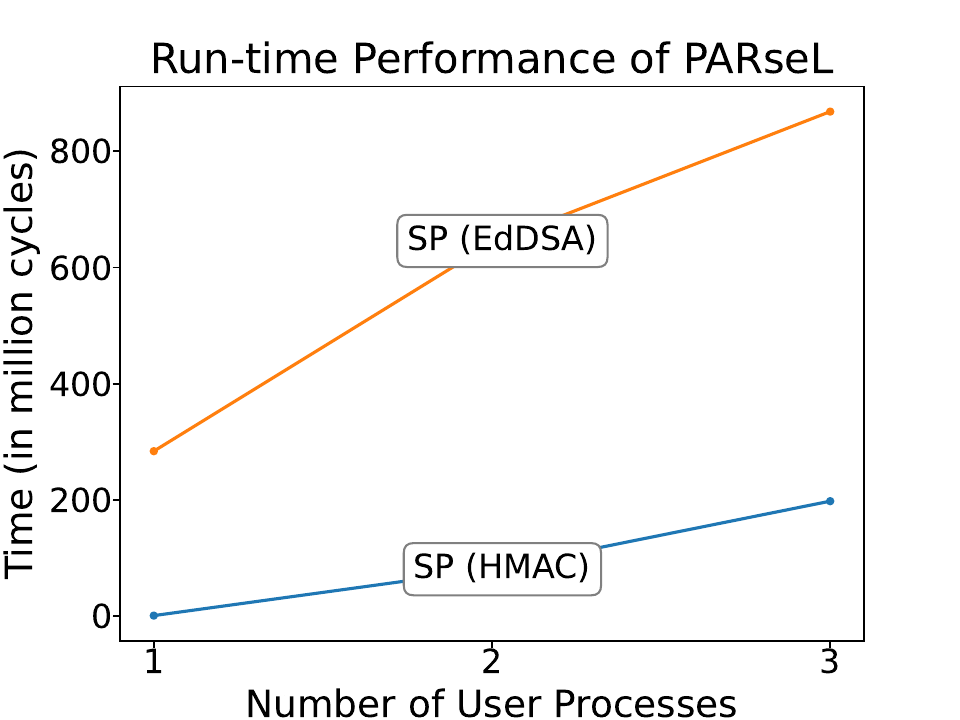}
	\caption{\footnotesize \fora Performance while varying the number of spawned user processes (excluding \SP)}
	\label{fig:eval}
 \vspace{-2pt}
\end{figure}

\subsection{Evaluation} \label{subsec:eval}

Our source code including verification proofs is available at \cite{parsel_code_anon}.

\subsubsection{Evaluation Setup}
To demonstrate the practicality of \fora, we developed our prototype on a commercially available hardware platform: SabreLite \cite{SabreLite}
-- on which \sel is fully verified~\cite{sel4_supported_platforms} including all proofs for functional correctness, integrity, and information flow.
SabreLite features an ARM Cortex-A9 CPU core (running at 1 GHz), with RAM of size 1 GB, and a microSD card slot (which we use to boot and load \fora image).
\fora is implemented on \sel version 12.0.1 (latest at the time of writing). 
Besides \sel IPC kernel APIs, \RP uses \sel Runtime, \sel Utils, and \sel Bench user-space libraries (offered by \sel Foundation) to implement \PST process spawning procedure. 

\subsubsection{\fora Performance}
The left sub-figure \revised{of Fig.~\ref{fig:eval}} shows the boot-time performance of \RP and \PST, and the right one shows the run-time performance of \SP (using either HMAC or EdDSA).
Reported results are averaged over 50 iterations. The size of each spawned process is $\approx0.4$ MB.

\RP takes constant 40 ms (40 million cycles @ 1 GHz), as it initiates the device and spawns \PST, independent of the number of \UP-s spawned. 
The time taken for \PST increases linearly to the number of \UP-s, as expected because \PST loads, measures, and spawns each \UP sequentially. 
Spawning each 0.4 MB \UP takes $\approx150$ ms.
Concretely, when there are 3 \UP-s, the boot-time of \fora is 1.3s. 

Using HMAC requires significantly fewer cycles than using EdDSA, due to its relatively expensive operations in the latter. 
\revised{The attestation time for one \UP using EdDSA} is 282 ms while \revised{the one} using HMAC is 1.2 ms.
As the number of \UP-s increase, the time taken for \SP also increases. 
This is due to frequent kernel context switching, as \sel (fully verified implementation) uses only one core.

\subsubsection{\fora TCB size}
\fora TCB contains $3.9$K lines of \Clang code, including 0.6K lines for \RP + \PST (excluding the \sel user-space libraries), and 3.3K lines for \SP.
Out of 3.3K lines of \SP, 3.2K lines are verified, including 3K lines from \hacl EdDSA and 0.2K lines from \SP run-time attestation function.

\section{\fora Security Analysis} \label{sec:security_analysis}

To argue \fora security with respect to the adversary model in Section~\ref{sec:adv_model}, we start by formulating \fora security goal.

\begin{mdframed}
\textit{
\noindent {\bf Security Definition:}
\begin{small}
Let $\mathcal{B}$ be an arbitrary software binary selected by \verifier.
In the context of a static root of trust for measurement of user-level processes, an \RA scheme is considered secure
if and only if \verifier is able to use the \RA scheme to establish a secure channel with program $\mathcal{P}$, where:\\
* $\mathcal{P}$ is an isolated user-level process running on the correct \prover;\\
* At boot time, $\mathcal{P}$ was loaded with the \verifier-selected binary $\mathcal{B}$;
\end{small}
}
\end{mdframed}

\noindent{\bf Security Argument:}
Assuming that \verifier uses \pk, included in $\sigma$ (recall Equation~\ref{eq:sign}), to establish the secure channel, \adv can attempt to circumvent \fora security by:

    \noindent{(1) \bf Loading the Right Software on the Wrong Device.} \adv can load process $P_{\adv}$ with the expected binary $\mathcal{B}$ on a different device ($\prover_{\adv}$), also equipped with an instance of \fora. Then, \adv forwards \verifier's request (intended to the original \prover) to $\prover_{\adv}$. $\prover_{\adv}$ inadvertently issues a \fora attestation response that matches software $\mathcal{B}$ (loaded on $P_{\adv}$). However, as the secret key $\key$ is unique to each \prover, \verifier would not accept the received $\sigma$, thereby refusing to establish the secure channel.

    \noindent{(2) \bf Loading the Wrong Software on the Right Device.} \adv can load a user-space process on the correct \prover but with an incorrect/malicious binary $\mathcal{B}_{\adv}$. This can be accomplished with physical access to \prover or by exploiting a vulnerability on a user-space process to perform persistent code injection, re-booting \prover thereafter. In either case, $\sigma$ would be signed with the expected secret key $\key$. However, \mmap would be updated at boot to reflect $\mathcal{B}_{\adv}$, i.e., the hash result $\texttt{m}_{\UP_{\adv}}$. Consequently, \verifier would refuse to establish a secure channel with a process on \prover loaded with $\mathcal{B}_{\adv}\neq\mathcal{B}$.

    \noindent{(3) \bf Loading the Wrong Software on the Wrong Device.} It follows from both arguments above that this option is infeasible to \adv due to the mismatches on both secret key $\key$ and measurement $\texttt{m}_{\UP_{\adv}}$.
    
%
%
Therefore, \fora satisfies the security definition above. 
$\qed$ 
\vspace{0.5em}

This argument assumes confidentiality of $\key$. In \fora, this is supported through formal verification of \SP functional correctness, secret independence, and memory safety.
It also assumes that each process is appropriately measured at boot. In \fora, this is implemented by \PST when computing \mmap. The association of \pk~ with the correct $\texttt{m}_{\UP}$ is guaranteed by \sel badge assignments. Finally, the scheme relies on inter-process isolation for \SP and any attested process $\mathcal{P}$, once the secure channel is established. The latter is inherited from \sel provable isolation.

\section{Discussion} \label{sec:limitations}
\noindent \textbf{Limitations:} 
Only \fora runtime TCB is verified. 
The integrity of \fora boot time TCB is ensured via secure boot, while the correct implementation of secure boot/boot TCB are assumed. 
Furthermore, \fora measures  
processes at boot time. Thus, \RP configures a write-xor-execution memory permission to prevent a user process from modifying its own code. By default, although \sel guarantees strong inter-process isolation, it gives each process full control of its own code/data segments. 
Due to this write-restriction, \fora does not support run-time updates to user-level processes.
Currently, benign updates must be done physically and require rebooting the device (in order to measure the updated program on boot).
However, we believe that any software update framework compatible with \sel (e.g. \cite{asokan2018assured}) can be used alongside with \fora for remote updates. 
The only requirement then would be to reboot the device after the update, so that \fora re-measures all \UP-s including the new updated \UP.

\noindent \textbf{(Unexpected) Termination of \UP} does not cause any issues because 
no other user process can transfer the signature (from \verifier) on behalf of another process to \SP. In \verifier's view, no response will arrive (in a certain amount of time) so it can deduce that \UP or \prover are no longer running.
This is similar to any \RA protocol.

\noindent \textbf{\SP Stack Erasure} is obviated in \fora because
\SP is never terminated at run-time and \sel's inter-process isolation guarantees that only \SP has access to its own stack.

\section{Related Work} \label{sec:related_work}

\textit{\RA:} techniques can be classified into SW-based, HW-based, and hybrid (HW/SW co-design) architectures. 
Although SW-based methods such as \cite{KeJa03,seshadri2004swatt,Viper2011,gligor} 
require minimal overall costs, 
they rely on strong assumptions about precise time-based checksum, 
which is mostly unsuitable for the IoT ecosystem with the multi-hop network. 
HW-based methods~\cite{sancus,tpm,KKW+12}, 
on the other hand, rely on some additional hardware support for \RA, 
e.g., some dedicated hardware components~\cite{tpm}, or extension of existing instruction sets~\cite{sgx}, which introduce cost and other barriers, especially for low-end and mid-range devices. 
Hybrid approach~\cite{vrasedp,trustlite,tytan,rata} 
is considered to be more suitable for IoT ecosystems because it aims for minimal hardware changes while keeping the same security levels as HW-based \RA.  
Using the hybrid \RA as a building block, many security services have been also suggested, such as proof of execution~\cite{apex,litehax}, control-flow and data-flow attestation~\cite{cflat,lofat,dialed,tinycfa,atrium,oat,geden2019hardware}, and secure software updates~\cite{asokan2018assured,pure,casu}.
Since \fora also provides a hybrid \RA, it can be also used for such security services.
%
Several recent papers on hybrid \RA/\RA-based security services~\cite{apex,vrasedp,pure,casu,rata} provide formal verification of their suggested architectures/implementations. 
They use model checking with temporal logic to verify their implementations while they use theorem proving to show that their proved properties are sufficient for their security goal(s). 

\textit{Verfied security applications in \fstar}:
\revised{\cite{papers_over_everest} lists papers that apply \fstar in security, including \hacl~\cite{hacl}.}
\dicestar~\cite{dicestar} is a notable paper related to \fora, which proposes a verified implementation of \textit{Device Identifier Composition Engine} (DICE), an industry-standard \textit{measured boot} protocol, for low-cost IoT devices. 
Similar to \fora, it has layered architecture with static components whose implementations are verified over \lowstar.
The main difference is how to guarantee the \key confidentiality.
DICE enforces the access control to the master secret key by locating it in a read-only and latchable memory so that only a hardware reset can disable/restore access to it.
The first hardware layer (called DICE engine) only has access to the secret, and it authenticates the next layer (L0) 
and derives the secret for L0 from its master secret and L0 measurement.
This ensures the same derived secret only when L0 firmware is not compromised.
\revised{Once received control, L0 derives a unique key pair from this secret and the next-layer firmware (L1); this key pair can then be used for L1 attestation and secure key exchange.}
Although \fora assumes a secure boot for correct \sel deployment, both \fora and \dicestar present verified implementations for the static root of trust for embedded devices, with different ways of guaranteeing the access control.

\textit{Architectures/applications over \sel}: 
After being released in 2009~\cite{seL4}, \sel has been actively implanted and used in both academia and industries in various domains, including automotive~\cite{GTTdemo}, aviation~\cite{CGB+18}, and medical devices~\cite{Russell20}. Apart from massive research from the Trustworthy Systems group in UNSW Sydney, many projects such as \cite{hydra,memocode21} leverage their architecture atop \sel.
%

\section{Conclusions}\label{sec:conclusion}

This paper presented \fora, a verifiable \RA root-of-trust over \sel.
We implemented it on SabreLite and demonstrated its overall feasibility and practicality. 
We also formally verified its runtime component in terms of functional correctness, memory safety, 
and secret independence, using the \lowstar tool-chain.
All source code, including verification proofs, is available at \cite{parsel_code_anon}.

\vspace{1em}
\noindent{\bf Acknowledgements:} We thank ICCAD'23 reviewers for constructive feedback. This work
was supported by funding from NSF Awards SATC-1956393, SATC-2245531,
and CICI-1840197, NSA Awards H98230-20-1-0345 and H98230-22-1-0308, 
as well as a subcontract from Peraton Labs. 
\newpage

\bibliographystyle{IEEEtran}

\bibliography{./references}
\end{document}